\documentclass[aps,prl,twocolumn,superscriptaddress,groupedaddress,showpacs]{revtex4}
\usepackage{graphicx}  
\usepackage{dcolumn}   
\usepackage{bm}        
\usepackage{amssymb}   

\usepackage{amsmath}
\usepackage{amssymb}
\usepackage{graphicx}
\usepackage{times}
\usepackage{rotating}
\usepackage{bm}
\usepackage{color}
\usepackage[breaklinks]{hyperref}
\renewcommand*{\HyperDestNameFilter}[1]{\jobname-#1}

\renewcommand{\vec}[1]{{\bf #1}}

\newcounter{lastnote}

\usepackage{float}
\begin{document}

\title{Dynamics and energetics of emergent magnetic monopoles in chiral magnets}

\author{Christoph Sch\"utte}
\affiliation{Institute for Theoretical Physics, University of Cologne, 50937 Cologne, Germany}
\author{Achim Rosch}
\affiliation{Institute for Theoretical Physics, University of Cologne, 50937 Cologne, Germany}

\date{\today}


\newcommand{\mfs}{Mn$_{1-x}$Fe$_{x}$Si}
\newcommand{\mcs}{Mn$_{1-x}$Co$_{x}$Si}
\newcommand{\fcs}{Fe$_{1-x}$Co$_{x}$Si}
\newcommand{\cso}{Cu$_{2}$OSeO$_{3}$}

\newcommand{\rxx}{$\rho_{xx}$}
\newcommand{\rxy}{$\rho_{xy}$}
\newcommand{\rxytop}{$\rho_{\rm xy}^{\rm top}$}
\newcommand{\Drxyt}{$\Delta\rho_{\rm xy}^{\rm top}$}
\newcommand{\Sxy}{$\sigma_{xy}$}
\newcommand{\Sxya}{$\sigma_{xy}^A$}

\newcommand{\bco}{$B_{\rm c1}$}
\newcommand{\bct}{$B_{\rm c2}$}
\newcommand{\bao}{$B_{\rm A1}$}
\newcommand{\bat}{$B_{\rm A2}$}
\newcommand{\beff}{$B_{\rm eff}$}

\newcommand{\tc}{$T_{\rm c}$}

\newcommand{\mb}{$\mu_0\,M/B$}
\newcommand{\dmdb}{$\mu_0\,\mathrm{d}M/\mathrm{d}B$}
\newcommand{\ddmddb}{$\mathrm{\mu_0\Delta}M/\mathrm{\Delta}B$}
\newcommand{\cm}{$\chi_{\rm M}$}
\newcommand{\cac}{$\chi_{\rm ac}$}
\newcommand{\rechi}{${\rm Re}\,\chi_{\rm ac}$}
\newcommand{\imchi}{${\rm Im}\,\chi_{\rm ac}$}

\newcommand{\ozz}{$\langle100\rangle$}
\newcommand{\ooz}{$\langle110\rangle$}
\newcommand{\ooo}{$\langle111\rangle$}
\newcommand{\too}{$\langle211\rangle$}

\renewcommand{\vec}[1]{{\bf #1}}


\begin{abstract}
The formation and destruction of topologically quantized magnetic whirls, so-called skyrmions, in chiral magnets is driven by the creation and motion of
singular hedgehog defects. These can be identified with emergent magnetic monopoles and antimonopoles.
We investigate how the energetics of and forces between monopoles and antimonopoles influence their
creation rate  and dynamics. We study a single skyrmion line defect in the helical phase using both micromagnetic simulations and a Ginzburg-Landau analysis. Monopole-antimonople pairs are created in a thermally activated process,
largely controlled by the (core) energy of the monopole. The force between monopoles and  antimonopoles
is linear in distance and described by a string tension.
The sign and size of the string tension determines the stability of the phases and the velocity of the monopoles.
\end{abstract}
 \maketitle

Small magnetic fields and thermal fluctuations can stabilize in chiral magnets lattices of magnetic whirls, so-called skyrmions \cite{muhlbauer2009skyrmion}. A skyrmion line is characterized by its topological property: the magnetization winds once around the unit-sphere for each plane cutting the line defect. This topological property is at the heart of a number of interesting properties of skyrmions \cite{fert2013skyrmions,schulz2012emergent,do2009skyrmions,zang2011dynamics,jonietz2010spin}. It is, for example, the main reason for a highly efficient coupling of the magnetic structure to electric currents: When an electron spin follows the local magnetization it  picks up a Berry phase. 
\begin{figure}[h]
  \centering
      \includegraphics[width=0.5\linewidth]{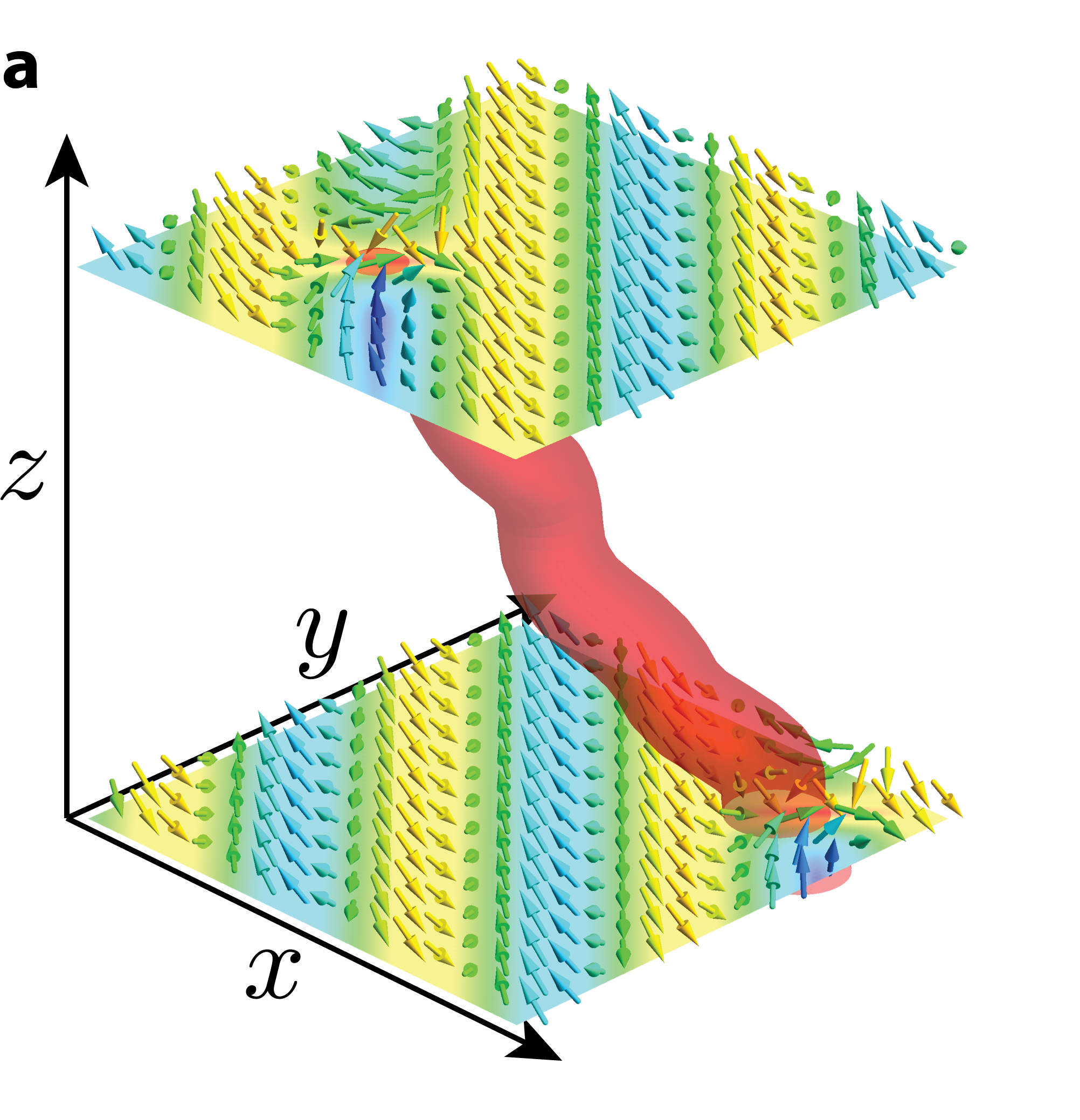}\\  \includegraphics[width=0.8\linewidth]{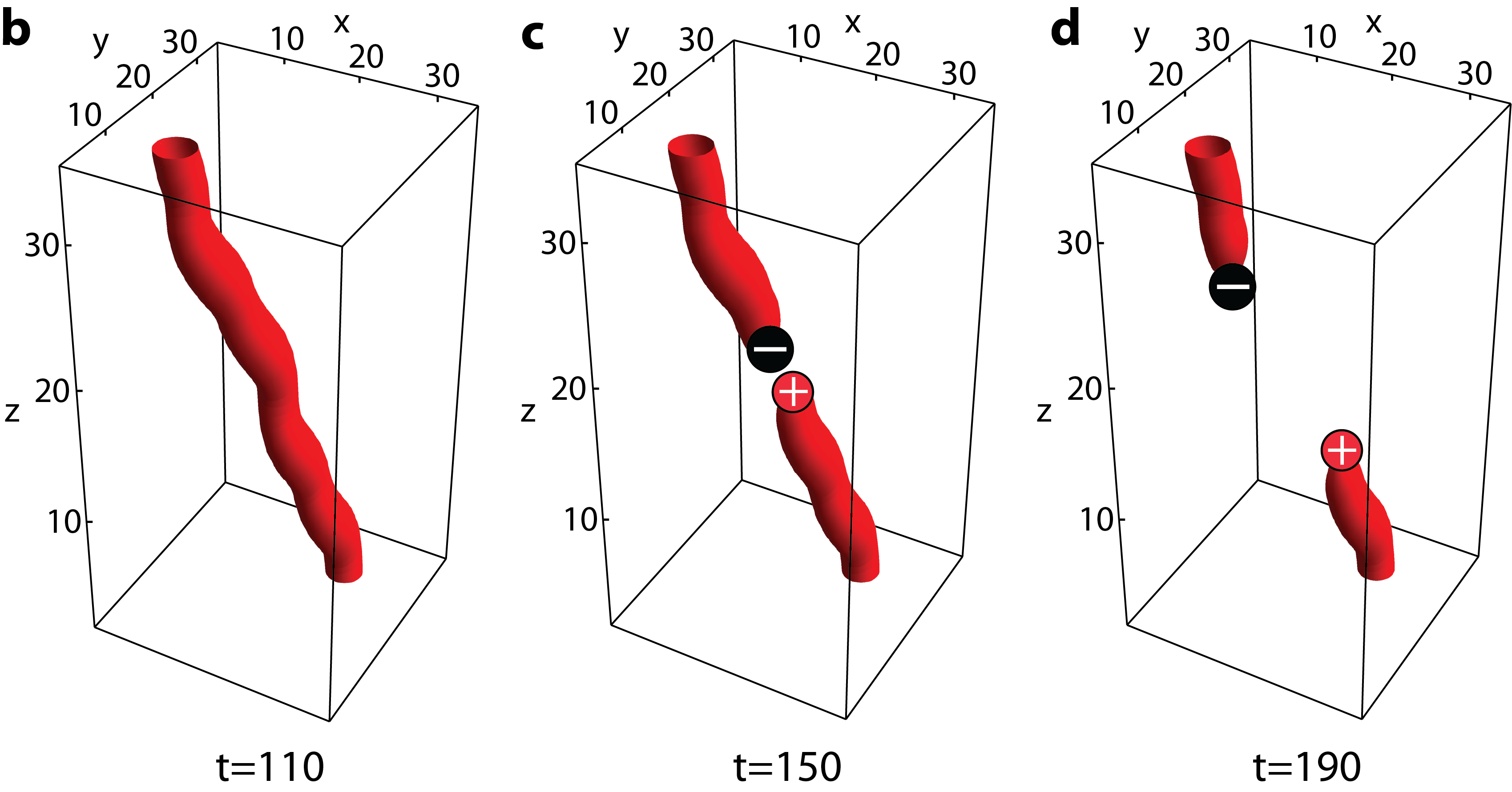} 
  \caption{(color online) a) Initial magnetic configuration: a skyrmion oriented along a (-1,-1,2) direction embedded into a helical phase. 
b-d) The skyrmion unwinds by the creation of a monopole-antimonopole pair which move apart (c.f. Fig.~\ref{fig2}a).\label{fig1}}
\end{figure}
This Berry phase can be described as an Aharonov-Bohm phase of an `artificial' electromagnetic fields. Each skyrmion line carries precisely one flux quantum of such an emergent magnetic field. The corresponding Lorentz force on the electron can be directly measured in Hall experiments and leads also to a Magnus force on the skyrmion in the presence of a current. Experimentally, one finds that skyrmions can be manipulated by extremely low electric currents, more than 5 orders of magnitude smaller than currents which are typically needed to move magnetic domain walls \cite{jonietz2010spin,schulz2012emergent}. 

Skyrmion lattices and single skyrmions can  be observed  by neutron scattering \cite{muhlbauer2009skyrmion,munzer2010skyrmion}, Lorentz-force electron microscopy \cite{yu2010near,yu2010real} or  magnetic force microscopy \cite{milde2013unwinding}.  With these techniques, not only regular lattices of skyrmion lines have been detected, but also single skyrmions and the crossover from the skyrmion phase to the helical phase. In the helical phase, the generic magnetic state
of chiral magnets for vanishing magnetic field, $B=0$, the magnetization rotates slowly perpendicular to a propagation vector $\vec q$.

Skyrmions have by now been observed in a wide range of chiral magnets, including good metals, doped semiconductors and even insulators. Using a different microscopic mechanism \cite{heinze2011spontaneous} tiny nanoskyrmions have also been stabilized in a hexagonal $\text{Fe}$ film  of one-atomic-layer thickness on a $\text{Ir}(111)$ substrate. In these systems the creation of single skyrmions can be triggered by a current passing through a magnetic tip of a scanning tunneling microscope \cite{heinze2011spontaneous,romming2013writing}.

As skyrmions are topologically stable, it is an interesting question how they can be created or destroyed. In a recent magnetic-force microscopy experiment \cite{milde2013unwinding}, it was shown that neighboring skyrmions merge when a transition from a skyrmion lattice phase to a phase with a simple helical order is induced by reducing the external magnetic field. A theoretical analysis and numerical simulations showed, that the destruction of skyrmions is driven by singular magnetic defects: The winding number of the skyrmions can only change at singular points where the magnetization vanishes. These singular points define hedgehog defects which act as sources and sinks of skyrmions and their associated magnetic fields. They can thereby considered as emergent magnetic monopoles (MPs) and antimonopoles (AMPs) \cite{milde2013unwinding}. In metals their motion induces
forces on the electrons which can be described by emergent electrodynamic fields \cite{schulz2012emergent,freimuth2013,takashima2014}.

In this paper, we study the creation and motion of MPs and AMPs. Motivated by the experimental setup in Ref. \cite{milde2013unwinding}, we investigate how skyrmions are destroyed and replaced by the helical phase. Previous numerical simulations \cite{milde2013unwinding} qualitatively investigated  this process starting from a dense skyrmion lattice. The simulations showed that the MP dynamics drives the transition. For a  quantitative analysis of this process, it is useful to simplify the problem by considering only the last step of the transition:
the destruction of the last skyrmion line. This helps to eliminate finite size effects in the numerical simulations:
 boundary effects have an influence on the energy difference of $N$ and $N-1$ skyrmions for large $N$.
Such problems are absent for a single skyrmion line, $N=1$, considered in the following. 

We use two different numerical methods, micromagnetic simulations based on the stochastic Landau Lifshitz Gilbert (sLLG) equation  and variational calculations based on a Ginzburg Landau description of helical magnets.
The stochastic Landau Lifshitz Gilbert equations (sLLG) \cite{garcia1998langevin} is given by
\begin{align}\label{sLLG}
\frac{d\mathbf{S_r}}{dt} = -\mathbf{S_r} \times \left[ \mathbf{B}_\mathbf{r}^{\text{eff}} + \mathbf{B}_\mathbf{r}^{\operatorname{fl}}(t) \right] + \alpha\,  \mathbf{S_r} \times \frac{d\mathbf{S_r}}{dt}
\end{align}
where the effective magnetic field $\mathbf{B}^{\text{eff}}_\mathbf{r}=-\frac{\delta H}{\delta \mathbf{S_r}}$ is obtained from the Hamiltonian $H$  
\begin{align}
H = &-J \sum_{\boldsymbol{r,\hat n =\hat x, \hat y,\hat z}} \boldsymbol{S}_{\boldsymbol{r}} \cdot  \boldsymbol{S}_{\boldsymbol{r}+{\boldsymbol{\hat{n}}}}  -  \boldsymbol{B}\cdot \sum_{\boldsymbol{r}} \boldsymbol{S}_{\boldsymbol{r}}\nonumber\\ 
			&-  K \sum_{\boldsymbol{r,\hat n =\hat x, \hat y,\hat z}}  \boldsymbol{S}_{\boldsymbol{r}} \times \boldsymbol{S}_{\boldsymbol{r}+{\boldsymbol{\hat{n}}}} \cdot \boldsymbol{\hat{n}} 
\label{eq:LatticeHamiltonian}
\end{align}
We use a cubic lattice and $K$ parametrizes the spin-orbit interactions ($K/J = \text{arctan}(2\pi/10)$ in our simulations).
The Gilbert damping $\alpha$ (we choose $\alpha=0.04$) describes spin relaxation. To simulate
the creation of MPs, it is essential to include the effects of thermal fluctuations which are described
by a randomly fluctuating magnetic field $\mathbf{B}^\text{fl}_\mathbf{r}(t)$ with
\begin{figure}[t]
  \centering
      \includegraphics[width=0.8\linewidth]{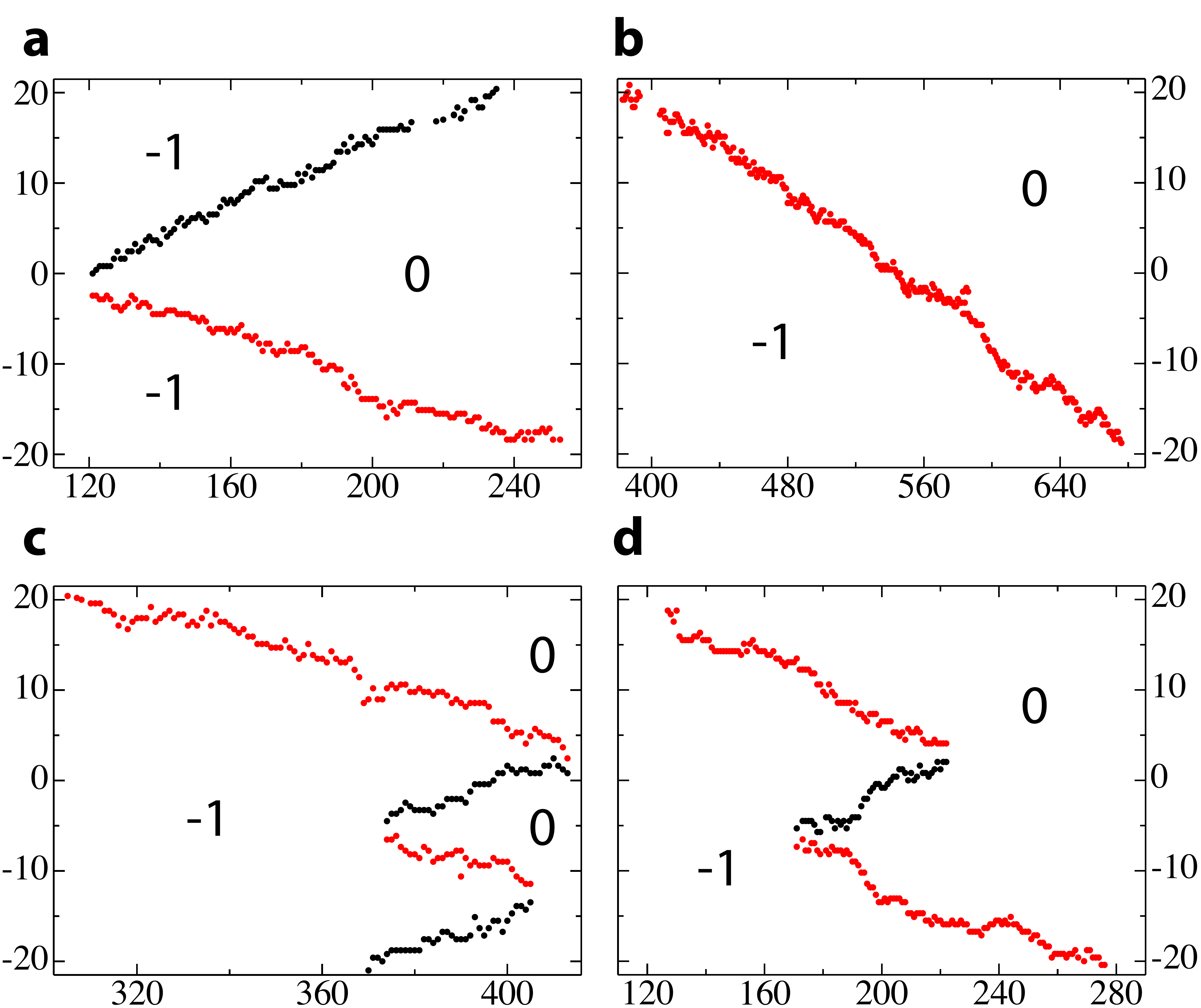}
  \caption{Examples of trajectories of MPs (black) and AMPs (red) after a quench to $B=0$ ($T=0.7$) obtained from sLLG simulations (vertical axis: coordinate parallel to the skyrmion orientation, numbers:  winding numbers of the magnetic texture) a) A MP-AMP pair is created in the middle of the sample and moves to the edge (c.f. Fig. ~\ref{fig1}). b) An AMP is created at the surface and moves to the bottom.  c, d) Events with both pair creation and pair annihilation.\label{fig2}}
\end{figure}
\begin{align}
&\langle \mathbf{B}^\text{fl}_\mathbf{r}(t) \rangle = 0, \quad
\langle B^\text{fl}_{i,\mathbf{r}}(t)B^\text{fl}_{j,\mathbf{r}}(t') = 2 \alpha k_B T \delta_{ij} \delta(t-t') 
\end{align}
consistent with the fluctuation-dissipation theorem, see Ref.~[\onlinecite{supplement}] for further implementation details.
The magnetic fields is applied in the $(-1,-1,2)$  direction, which is chosen perpendicular to the $(111)$ direction to avoid a tilting of helical phase which has an ordering vector in $(111)$ direction.
 At $t=0$ we start from a configuration described by a helical
spin state which has a single skyrmion embedded, see Fig.~\ref{fig1}a. 

In Fig.~\ref{fig1}b-d the red line tracks the center of mass, $\int \vec r \rho_t dx dy$, of the skyrmion where
$\rho_t=\frac{1}{4 \pi} \hat M\cdot (\partial_x \hat M \times \partial_y \hat M)$ is the topological
charge density. The three figures show an event (also displayed in Fig.~\ref{fig2}a), where the  
skyrmion string is cut into two by the creation of a MP-AMP pair: As the winding number $\int \rho_t dx dy$ changes at the end of
the skyrmion string from $-1$ to $0$, there has to be a topological defect (a hedgehog) at the end of the string. The hedgehog with  winding number $+1$ ($-1$) can be viewed \cite{milde2013unwinding} as a magnetic MP (AMP) and is denoted by $-$ sign ($+$ sign) in Fig.~\ref{fig1}, respectively. 

In Fig.~\ref{fig2} we show typical trajectories of MPs and AMPs obtained from sLLG simulations after the magnetic field has been switched to $0$. For $B=0$ the skyrmion state is unstable,
therefore MPs and AMPs are created spontaneously by thermal fluctuations.
 As the MPs move predominantly parallel to the skyrmion orientation, we show on the vertical axis the projection of the  MP coordinate onto the $(-1,-1,2)$ axis. In the bulk, MPs and AMPs are always created as pairs due to their topology. MPs move `up', AMPs `down' to reduce the winding number (see Fig.~\ref{fig1}).
Single MPs or AMPs can only be created at the top or bottom layer (Fig.~\ref{fig2}b,c,d). 
Fig.~\ref{fig2}c and d show that also MP and AMP annihilate each other when they 
come close to each other.

\begin{figure}[t]
  \centering
  \includegraphics[width=0.99\linewidth]{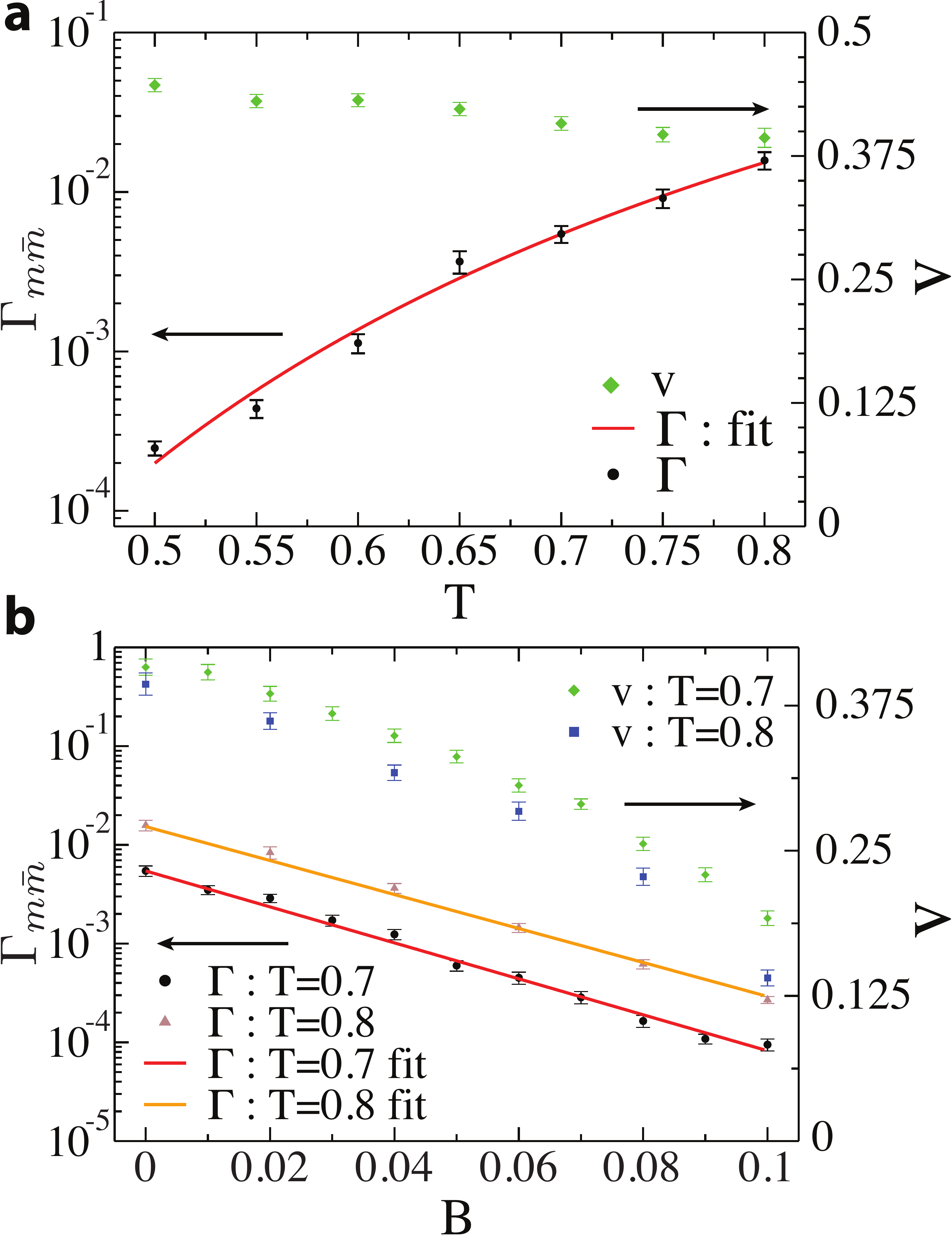}
  \caption{(a) Average MP velocity, $v$ (right legend), and average bulk creation rate, $\Gamma$ (left legend), as a function of the temperature $T$ for $B=0$ (error bars: standard deviation of the mean)\label{fig3}. (b) $v$ and $\Gamma$ as a function of the applied magnetic field $B$ for
 $T=0.7$ and $T=0.8$. }
\end{figure}
To describe the dynamics of MPs and AMPs two quantities are of main interest:
the creation rate $\Gamma_{m\bar m}$ of MP-AMP pairs (defined per length of the skyrmion line) and the average velocity of MPs. In the supplementary material we also discuss briefly the creation rate $\Gamma_m^s$ of single MPs at the surface of the sample.
In Fig.~\ref{fig3}, $\Gamma_{m\bar m}$ is plotted as a function of $T$. The fit shows that the  $T$ dependence is consistent with a simple activated behavior
\begin{eqnarray}
\Gamma_{m\bar m}\sim \Gamma_0 e^{-\frac{E_0}{k_B T}}\label{exp}
\end{eqnarray}
with $E_0 \approx 5.8\,J$ for the chosen parameters. The creation rate of MPs also strongly
depends on the magnetic field, see Fig..~\ref{fig3}b, $\Gamma_{m\bar m}\propto e^{-B/B_0}$. A comparison of simulations at  $T=0.7$ and $T=0.8$ (not shown) seems to suggest that the exponential dependence arises from a combination of the $B$ dependence of the activation energy $E_0$ and of the prefactor $\Gamma_0$ but a definite conclusion is not possible from the available data. 
The average velocity of the MPs depends only weakly on $T$ but as a function of magnetic field, it is suppressed by a factor of about $3$ in the considered  $B$ range, see Fig.~\ref{fig3}a and b.

\begin{figure}[t]
  \centering
      \includegraphics[width=0.8\linewidth]{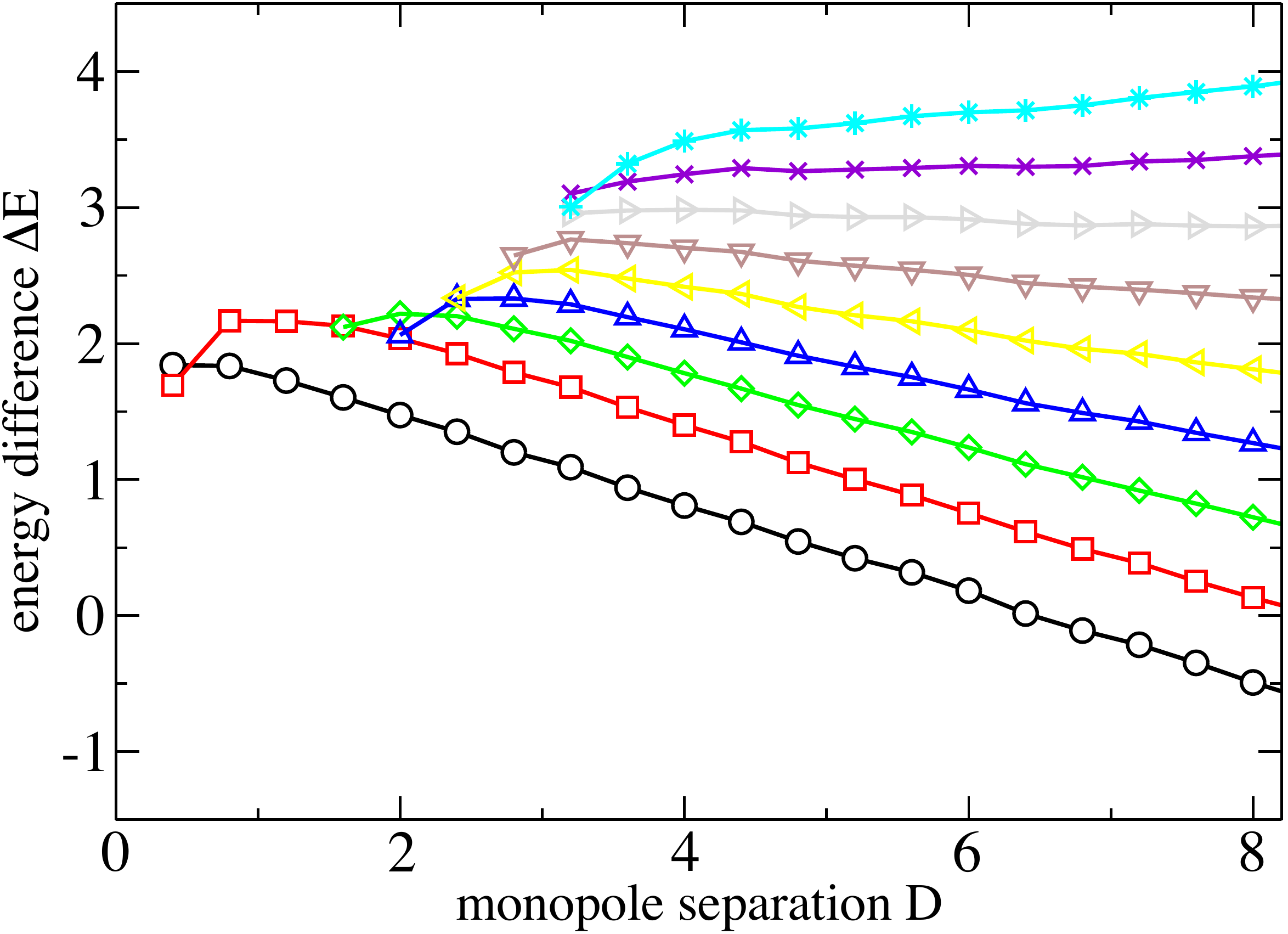}
  \caption{Energy of a MP-AMP pair as a function of their distance obtained from minimization of the GL free energy ($t_0 =- 1.6$ \cite{supplement}) for $B=0.14$ to $0.22$ (bottom to top). For large distances (here we use units where the pitch of the helical phase is $2 \pi$ \cite{supplement}) the force is linear and therefore described by a string tension (see Fig.~\ref{fig5})\label{fig4}}
\end{figure}
To explain the results of the sLLG simulations, we consider the energetics of MP-AMP pairs. It is mainly determined by two factors: the free energy needed locally to create the singular magnetic configuration of a MP (the {\em core energy}, $E_c$)  and the free energy per length of a skyrmion.
The latter gives rise to a {\em string tension}, $T_S$: 
the skyrmion pulls at the MP with a constant force, $F=T_S$. The system gains the energy
$T_S \Delta x$ when the MP moves the distance $\Delta x$, thus shortening the length of the skyrmion string. The string tension can also be viewed as resulting from an interaction potential of MPs and AMPs linear in distance, see below.

\begin{figure}[t]
  \centering
  \includegraphics[width=0.9\linewidth]{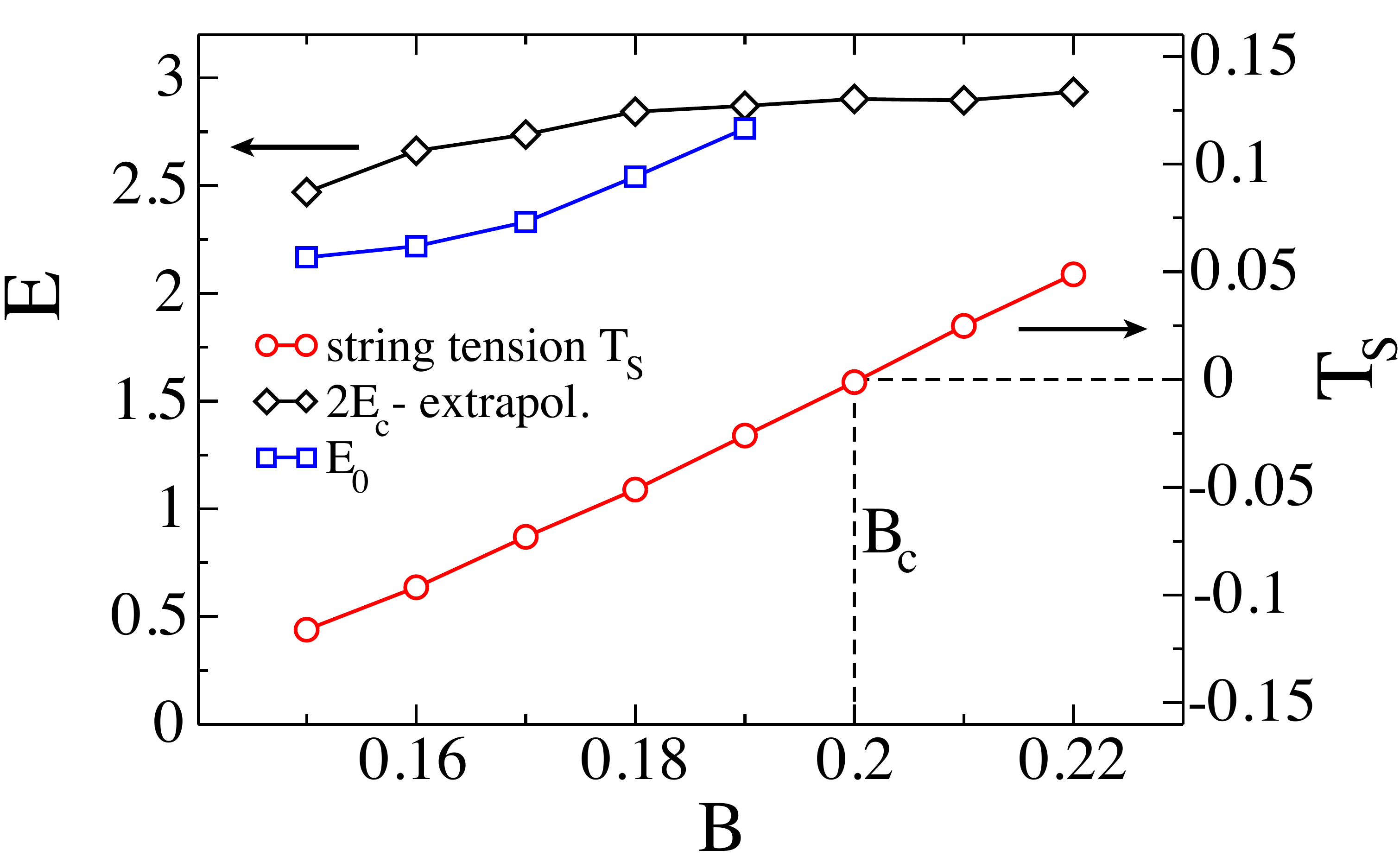}
  \caption{Energy $E_0$ of a MP-AMP pair, sum of their core energies, $2 E_c$ (both left axis), and  string tension $T_S$ (right axis) as a function of the magnetic field $\mathbf B_z$ ($t_0=-1.6$, see \cite{supplement}). At the dashed line the string tension changes sign.\label{fig5}}
\end{figure}
To corroborate this physical picture using an alternative theoretical approach, we have calculated the energetics of skyrmions and MP-AMP pairs using a Ginzburg Landau (GL) description of the free energy. We use the standard Ginzburg Landau functional for chiral magnets \cite{bak1980theory} discretized on a 50$\times$50$\times 50$ lattice, see Ref.~[\onlinecite{supplement}] for details.
Our main goal is to obtain the free energy of a MP-AMP pair as a function of their distance $D$. To fix $D$ we use that in the continuum the magnetization $M$ vanishes in the core of each MP and AMP. We therefore fix their positions by demanding that
$M=0$ at two lattice sites with distance $D$. It turns out that this procedure works 
only as long as the forces on the MPs and AMPs are not stronger than the pinning energy due to
setting $M=0$ at one site. This limits our GL study to a finite field range, parameters not too far \cite{supplement} from $T_c$, and not too small distances $D$. A typical result is shown in Fig.~\ref{fig4}.
For large distances the energy is linear in $D$. The slope is - by definition - the string tension $T_S$, shown in Fig.~\ref{fig5} as function of magnetic field. 
As expected, $T_S$ is exactly given by the energy per length of a skyrmion. As a function of $B$, $T_S$ changes its sign at a critical value $B_c$. For $B<B_c$ the skyrmion is not stable, $T_S>0$ and MP and AMP repel each other at large distances. For $B>B_c$ a single skyrmion has a lower energy than the
helical state, $T_S<0$, and the MP-AMP interaction is attractive for long distances. We have checked that in our sLLG simulation (not shown) this leads to the spontaneous creation of skyrmions in this regime. Note that the bulk phase transition is not given by $B_c$ due to skyrmion-skyrmion interactions in a dense skyrmion lattice.

For $T_S>0$, the MP-AMP energy is always negative for sufficiently large  $D$. The creation rate is, however, strongly suppressed by the fact that for small $D$ a MP-AMP pair has a large, positive energy.  
The main contribution to this energy is the core energy, $E_c$, of a single MP or AMP which can be obtained by a linear extrapolation of $\Delta E$ in Fig.~\ref{fig4} to $D=0$, which gives
$2 E_c$. The figure shows, however, that there is a short range attraction of the MP-AMP pair. Therefore we expect that instead of $2 E_c$, the maximum $E_0=\max_D \Delta E$ controls the MP-AMP creation rate of Eq.~\ref{exp}. $2 E_c$ and $E_0$ are plotted for comparison in Fig.~\ref{fig5}.

A quantitative comparison of the sLLG simulation and the GL calculation was not possible in our study as the range of parameters where each method can be applied, did not overlap. One can, however, compare the results qualitatively. First, the velocity of MPs is expected to be given by the product of the string tension $T_s$ and an effective friction coefficient. Assuming approximately constant friction, we expect from the $B$ (Fig.~\ref{fig5}) and $T$ (see supplement \cite{supplement}) dependence of $T_s$ that the MP velocity drops with $B$ for $B>B_c$ and rises when lowering $T$ as indeed observed in the sLLG simulations, see Fig.~\ref{fig3}. More dramatically, an increase of the core energy for increasing $B$
 should give rise to an exponential decrease of the MP-AMP creation rate $\sim e^{-E_c/k_B T}$ as a function of $B$, as observed numerically in Fig.~\ref{fig3}a.

In conclusion,  string tension and core energy of monopoles are the dominating factors which determine the creation rates and the dynamics of monopoles. These quantities are the key to understand how
skyrmion lines (and lattices thereof) can be created and destroyed in three-dimensional bulk materials. A related question -- especially relevant for future skyrmion-based devices -- is, how skyrmions in two-dimensional films can be
created and destroyed \cite{romming2013writing,iwasaki2013current,sampaio2013nucleation}. 
When one replaces the $z$ axis, e.g. in Fig.~\ref{fig1}, by a time axis,
one realizes that a monopole spin-configuration can be viewed as an instanton describing the
destruction or creation of a two-dimensional skyrmion. We expect that the corresponding instanton 
action which controls the creation rates, will be dominated by a contribution related (but not identical) to the core energy of the monopole.

\bibliographystyle{apsrev}
\bibliography{bibliography}

\begin{thebibliography}{19}
\expandafter\ifx\csname natexlab\endcsname\relax\def\natexlab#1{#1}\fi
\expandafter\ifx\csname bibnamefont\endcsname\relax
  \def\bibnamefont#1{#1}\fi
\expandafter\ifx\csname bibfnamefont\endcsname\relax
  \def\bibfnamefont#1{#1}\fi
\expandafter\ifx\csname citenamefont\endcsname\relax
  \def\citenamefont#1{#1}\fi
\expandafter\ifx\csname url\endcsname\relax
  \def\url#1{\texttt{#1}}\fi
\expandafter\ifx\csname urlprefix\endcsname\relax\def\urlprefix{URL }\fi
\providecommand{\bibinfo}[2]{#2}
\providecommand{\eprint}[2][]{\url{#2}}

\bibitem[{\citenamefont{M{\"u}hlbauer et~al.}(2009)\citenamefont{M{\"u}hlbauer,
  Binz, Jonietz, Pfleiderer, Rosch, Neubauer, Georgii, and
  B{\"o}ni}}]{muhlbauer2009skyrmion}
\bibinfo{author}{\bibfnamefont{S.}~\bibnamefont{M{\"u}hlbauer}},
  \bibinfo{author}{\bibfnamefont{B.}~\bibnamefont{Binz}},
  \bibinfo{author}{\bibfnamefont{F.}~\bibnamefont{Jonietz}},
  \bibinfo{author}{\bibfnamefont{C.}~\bibnamefont{Pfleiderer}},
  \bibinfo{author}{\bibfnamefont{A.}~\bibnamefont{Rosch}},
  \bibinfo{author}{\bibfnamefont{A.}~\bibnamefont{Neubauer}},
  \bibinfo{author}{\bibfnamefont{R.}~\bibnamefont{Georgii}}, \bibnamefont{and}
  \bibinfo{author}{\bibfnamefont{P.}~\bibnamefont{B{\"o}ni}},
  \bibinfo{journal}{Science} \textbf{\bibinfo{volume}{323}},
  \bibinfo{pages}{915} (\bibinfo{year}{2009}).

\bibitem[{\citenamefont{Fert et~al.}(2013)\citenamefont{Fert, Cros, and
  Sampaio}}]{fert2013skyrmions}
\bibinfo{author}{\bibfnamefont{A.}~\bibnamefont{Fert}},
  \bibinfo{author}{\bibfnamefont{V.}~\bibnamefont{Cros}}, \bibnamefont{and}
  \bibinfo{author}{\bibfnamefont{J.}~\bibnamefont{Sampaio}},
  \bibinfo{journal}{Nature Nanotechnology} \textbf{\bibinfo{volume}{8}},
  \bibinfo{pages}{152} (\bibinfo{year}{2013}).

\bibitem[{\citenamefont{Schulz et~al.}(2012)\citenamefont{Schulz, Ritz, Bauer,
  Halder, Wagner, Franz, Pfleiderer, Everschor, Garst, and
  Rosch}}]{schulz2012emergent}
\bibinfo{author}{\bibfnamefont{T.}~\bibnamefont{Schulz}},
  \bibinfo{author}{\bibfnamefont{R.}~\bibnamefont{Ritz}},
  \bibinfo{author}{\bibfnamefont{A.}~\bibnamefont{Bauer}},
  \bibinfo{author}{\bibfnamefont{M.}~\bibnamefont{Halder}},
  \bibinfo{author}{\bibfnamefont{M.}~\bibnamefont{Wagner}},
  \bibinfo{author}{\bibfnamefont{C.}~\bibnamefont{Franz}},
  \bibinfo{author}{\bibfnamefont{C.}~\bibnamefont{Pfleiderer}},
  \bibinfo{author}{\bibfnamefont{K.}~\bibnamefont{Everschor}},
  \bibinfo{author}{\bibfnamefont{M.}~\bibnamefont{Garst}}, \bibnamefont{and}
  \bibinfo{author}{\bibfnamefont{A.}~\bibnamefont{Rosch}},
  \bibinfo{journal}{Nature Physics} \textbf{\bibinfo{volume}{8}},
  \bibinfo{pages}{301} (\bibinfo{year}{2012}).

\bibitem[{\citenamefont{Do~Yi et~al.}(2009)\citenamefont{Do~Yi, Onoda, Nagaosa,
  and Han}}]{do2009skyrmions}
\bibinfo{author}{\bibfnamefont{S.}~\bibnamefont{Do~Yi}},
  \bibinfo{author}{\bibfnamefont{S.}~\bibnamefont{Onoda}},
  \bibinfo{author}{\bibfnamefont{N.}~\bibnamefont{Nagaosa}}, \bibnamefont{and}
  \bibinfo{author}{\bibfnamefont{J.~H.} \bibnamefont{Han}},
  \bibinfo{journal}{Physical Review B} \textbf{\bibinfo{volume}{80}},
  \bibinfo{pages}{054416} (\bibinfo{year}{2009}).

\bibitem[{\citenamefont{Zang et~al.}(2011)\citenamefont{Zang, Mostovoy, Han,
  and Nagaosa}}]{zang2011dynamics}
\bibinfo{author}{\bibfnamefont{J.}~\bibnamefont{Zang}},
  \bibinfo{author}{\bibfnamefont{M.}~\bibnamefont{Mostovoy}},
  \bibinfo{author}{\bibfnamefont{J.~H.} \bibnamefont{Han}}, \bibnamefont{and}
  \bibinfo{author}{\bibfnamefont{N.}~\bibnamefont{Nagaosa}},
  \bibinfo{journal}{Physical review letters} \textbf{\bibinfo{volume}{107}},
  \bibinfo{pages}{136804} (\bibinfo{year}{2011}).

\bibitem[{\citenamefont{Jonietz et~al.}(2010)\citenamefont{Jonietz,
  M\"uhlbauer, Pfleiderer, Neubauer, M\"unzer, Bauer, Adams, Georgii, B\"oni,
  Duine et~al.}}]{jonietz2010spin}
\bibinfo{author}{\bibfnamefont{F.}~\bibnamefont{Jonietz}},
  \bibinfo{author}{\bibfnamefont{S.}~\bibnamefont{M\"uhlbauer}},
  \bibinfo{author}{\bibfnamefont{C.}~\bibnamefont{Pfleiderer}},
  \bibinfo{author}{\bibfnamefont{A.}~\bibnamefont{Neubauer}},
  \bibinfo{author}{\bibfnamefont{W.}~\bibnamefont{M\"unzer}},
  \bibinfo{author}{\bibfnamefont{A.}~\bibnamefont{Bauer}},
  \bibinfo{author}{\bibfnamefont{T.}~\bibnamefont{Adams}},
  \bibinfo{author}{\bibfnamefont{R.}~\bibnamefont{Georgii}},
  \bibinfo{author}{\bibfnamefont{P.}~\bibnamefont{B\"oni}},
  \bibinfo{author}{\bibfnamefont{R.~A.} \bibnamefont{Duine}},
  \bibnamefont{et~al.}, \bibinfo{journal}{Science}
  \textbf{\bibinfo{volume}{330}}, \bibinfo{pages}{1648} (\bibinfo{year}{2010}).

\bibitem[{\citenamefont{M{\"u}nzer et~al.}(2010)\citenamefont{M{\"u}nzer,
  Neubauer, Adams, M{\"u}hlbauer, Franz, Jonietz, Georgii, B{\"o}ni, Pedersen,
  Schmidt et~al.}}]{munzer2010skyrmion}
\bibinfo{author}{\bibfnamefont{W.}~\bibnamefont{M{\"u}nzer}},
  \bibinfo{author}{\bibfnamefont{A.}~\bibnamefont{Neubauer}},
  \bibinfo{author}{\bibfnamefont{T.}~\bibnamefont{Adams}},
  \bibinfo{author}{\bibfnamefont{S.}~\bibnamefont{M{\"u}hlbauer}},
  \bibinfo{author}{\bibfnamefont{C.}~\bibnamefont{Franz}},
  \bibinfo{author}{\bibfnamefont{F.}~\bibnamefont{Jonietz}},
  \bibinfo{author}{\bibfnamefont{R.}~\bibnamefont{Georgii}},
  \bibinfo{author}{\bibfnamefont{P.}~\bibnamefont{B{\"o}ni}},
  \bibinfo{author}{\bibfnamefont{B.}~\bibnamefont{Pedersen}},
  \bibinfo{author}{\bibfnamefont{M.}~\bibnamefont{Schmidt}},
  \bibnamefont{et~al.}, \bibinfo{journal}{Physical Review B}
  \textbf{\bibinfo{volume}{81}}, \bibinfo{pages}{041203}
  (\bibinfo{year}{2010}).

\bibitem[{\citenamefont{Yu et~al.}(2010{\natexlab{a}})\citenamefont{Yu,
  Kanazawa, Onose, Kimoto, Zhang, Ishiwata, Matsui, and Tokura}}]{yu2010near}
\bibinfo{author}{\bibfnamefont{X.}~\bibnamefont{Yu}},
  \bibinfo{author}{\bibfnamefont{N.}~\bibnamefont{Kanazawa}},
  \bibinfo{author}{\bibfnamefont{Y.}~\bibnamefont{Onose}},
  \bibinfo{author}{\bibfnamefont{K.}~\bibnamefont{Kimoto}},
  \bibinfo{author}{\bibfnamefont{W.}~\bibnamefont{Zhang}},
  \bibinfo{author}{\bibfnamefont{S.}~\bibnamefont{Ishiwata}},
  \bibinfo{author}{\bibfnamefont{Y.}~\bibnamefont{Matsui}}, \bibnamefont{and}
  \bibinfo{author}{\bibfnamefont{Y.}~\bibnamefont{Tokura}},
  \bibinfo{journal}{Nature materials} \textbf{\bibinfo{volume}{10}},
  \bibinfo{pages}{106} (\bibinfo{year}{2010}{\natexlab{a}}).

\bibitem[{\citenamefont{Yu et~al.}(2010{\natexlab{b}})\citenamefont{Yu, Onose,
  Kanazawa, Park, Han, Matsui, Nagaosa, and Tokura}}]{yu2010real}
\bibinfo{author}{\bibfnamefont{X.}~\bibnamefont{Yu}},
  \bibinfo{author}{\bibfnamefont{Y.}~\bibnamefont{Onose}},
  \bibinfo{author}{\bibfnamefont{N.}~\bibnamefont{Kanazawa}},
  \bibinfo{author}{\bibfnamefont{J.}~\bibnamefont{Park}},
  \bibinfo{author}{\bibfnamefont{J.}~\bibnamefont{Han}},
  \bibinfo{author}{\bibfnamefont{Y.}~\bibnamefont{Matsui}},
  \bibinfo{author}{\bibfnamefont{N.}~\bibnamefont{Nagaosa}}, \bibnamefont{and}
  \bibinfo{author}{\bibfnamefont{Y.}~\bibnamefont{Tokura}},
  \bibinfo{journal}{Nature} \textbf{\bibinfo{volume}{465}},
  \bibinfo{pages}{901} (\bibinfo{year}{2010}{\natexlab{b}}).

\bibitem[{\citenamefont{Milde et~al.}(2013)\citenamefont{Milde, K{\"o}hler,
  Seidel, Eng, Bauer, Chacon, Kindervater, M{\"u}hlbauer, Pfleiderer, Buhrandt
  et~al.}}]{milde2013unwinding}
\bibinfo{author}{\bibfnamefont{P.}~\bibnamefont{Milde}},
  \bibinfo{author}{\bibfnamefont{D.}~\bibnamefont{K{\"o}hler}},
  \bibinfo{author}{\bibfnamefont{J.}~\bibnamefont{Seidel}},
  \bibinfo{author}{\bibfnamefont{L.}~\bibnamefont{Eng}},
  \bibinfo{author}{\bibfnamefont{A.}~\bibnamefont{Bauer}},
  \bibinfo{author}{\bibfnamefont{A.}~\bibnamefont{Chacon}},
  \bibinfo{author}{\bibfnamefont{J.}~\bibnamefont{Kindervater}},
  \bibinfo{author}{\bibfnamefont{S.}~\bibnamefont{M{\"u}hlbauer}},
  \bibinfo{author}{\bibfnamefont{C.}~\bibnamefont{Pfleiderer}},
  \bibinfo{author}{\bibfnamefont{S.}~\bibnamefont{Buhrandt}},
  \bibnamefont{et~al.}, \bibinfo{journal}{Science}
  \textbf{\bibinfo{volume}{340}}, \bibinfo{pages}{1076} (\bibinfo{year}{2013}).

\bibitem[{\citenamefont{Heinze et~al.}(2011)\citenamefont{Heinze, von Bergmann,
  Menzel, Brede, Kubetzka, Wiesendanger, Bihlmayer, and
  Bl{\"u}gel}}]{heinze2011spontaneous}
\bibinfo{author}{\bibfnamefont{S.}~\bibnamefont{Heinze}},
  \bibinfo{author}{\bibfnamefont{K.}~\bibnamefont{von Bergmann}},
  \bibinfo{author}{\bibfnamefont{M.}~\bibnamefont{Menzel}},
  \bibinfo{author}{\bibfnamefont{J.}~\bibnamefont{Brede}},
  \bibinfo{author}{\bibfnamefont{A.}~\bibnamefont{Kubetzka}},
  \bibinfo{author}{\bibfnamefont{R.}~\bibnamefont{Wiesendanger}},
  \bibinfo{author}{\bibfnamefont{G.}~\bibnamefont{Bihlmayer}},
  \bibnamefont{and}
  \bibinfo{author}{\bibfnamefont{S.}~\bibnamefont{Bl{\"u}gel}},
  \bibinfo{journal}{Nature Physics} \textbf{\bibinfo{volume}{7}},
  \bibinfo{pages}{713} (\bibinfo{year}{2011}).

\bibitem[{\citenamefont{Romming et~al.}(2013)\citenamefont{Romming, Hanneken,
  Menzel, Bickel, Wolter, von Bergmann, Kubetzka, and
  Wiesendanger}}]{romming2013writing}
\bibinfo{author}{\bibfnamefont{N.}~\bibnamefont{Romming}},
  \bibinfo{author}{\bibfnamefont{C.}~\bibnamefont{Hanneken}},
  \bibinfo{author}{\bibfnamefont{M.}~\bibnamefont{Menzel}},
  \bibinfo{author}{\bibfnamefont{J.~E.} \bibnamefont{Bickel}},
  \bibinfo{author}{\bibfnamefont{B.}~\bibnamefont{Wolter}},
  \bibinfo{author}{\bibfnamefont{K.}~\bibnamefont{von Bergmann}},
  \bibinfo{author}{\bibfnamefont{A.}~\bibnamefont{Kubetzka}}, \bibnamefont{and}
  \bibinfo{author}{\bibfnamefont{R.}~\bibnamefont{Wiesendanger}},
  \bibinfo{journal}{Science} \textbf{\bibinfo{volume}{341}},
  \bibinfo{pages}{636} (\bibinfo{year}{2013}).

\bibitem[{\citenamefont{Freimuth et~al.}(2013)\citenamefont{Freimuth, Bamler,
  Mokrousov, and Rosch}}]{freimuth2013}
\bibinfo{author}{\bibfnamefont{F.}~\bibnamefont{Freimuth}},
  \bibinfo{author}{\bibfnamefont{R.}~\bibnamefont{Bamler}},
  \bibinfo{author}{\bibfnamefont{Y.}~\bibnamefont{Mokrousov}},
  \bibnamefont{and} \bibinfo{author}{\bibfnamefont{A.}~\bibnamefont{Rosch}},
  \bibinfo{journal}{Phys. Rev. B} \textbf{\bibinfo{volume}{88}},
  \bibinfo{pages}{214409} (\bibinfo{year}{2013}).

\bibitem[{\citenamefont{{Takashima} and {Fujimoto}}(2014)}]{takashima2014}
\bibinfo{author}{\bibfnamefont{R.}~\bibnamefont{{Takashima}}} \bibnamefont{and}
  \bibinfo{author}{\bibfnamefont{S.}~\bibnamefont{{Fujimoto}}},
  \bibinfo{journal}{ArXiv e-prints}  (\bibinfo{year}{2014}),
  \eprint{1401.7140}.

\bibitem[{\citenamefont{Garc{\'\i}a-Palacios and
  L{\'a}zaro}(1998)}]{garcia1998langevin}
\bibinfo{author}{\bibfnamefont{J.~L.} \bibnamefont{Garc{\'\i}a-Palacios}}
  \bibnamefont{and} \bibinfo{author}{\bibfnamefont{F.~J.}
  \bibnamefont{L{\'a}zaro}}, \bibinfo{journal}{Physical Review B}
  \textbf{\bibinfo{volume}{58}}, \bibinfo{pages}{14937} (\bibinfo{year}{1998}).

\bibitem[{\citenamefont{supplementary material}()}]{supplement}
\bibinfo{author}{\bibfnamefont{S.}~\bibnamefont{supplementary material}}
  (????).

\bibitem[{\citenamefont{Bak and Jensen}(1980)}]{bak1980theory}
\bibinfo{author}{\bibfnamefont{P.}~\bibnamefont{Bak}} \bibnamefont{and}
  \bibinfo{author}{\bibfnamefont{M.~H.} \bibnamefont{Jensen}},
  \bibinfo{journal}{Journal of Physics C: Solid State Physics}
  \textbf{\bibinfo{volume}{13}}, \bibinfo{pages}{L881} (\bibinfo{year}{1980}).

\bibitem[{\citenamefont{Iwasaki et~al.}(2013)\citenamefont{Iwasaki, Mochizuki,
  and Nagaosa}}]{iwasaki2013current}
\bibinfo{author}{\bibfnamefont{J.}~\bibnamefont{Iwasaki}},
  \bibinfo{author}{\bibfnamefont{M.}~\bibnamefont{Mochizuki}},
  \bibnamefont{and} \bibinfo{author}{\bibfnamefont{N.}~\bibnamefont{Nagaosa}},
  \bibinfo{journal}{Nature nanotechnology}  (\bibinfo{year}{2013}).

\bibitem[{\citenamefont{Sampaio et~al.}(2013)\citenamefont{Sampaio, Cros,
  Rohart, Thiaville, and Fert}}]{sampaio2013nucleation}
\bibinfo{author}{\bibfnamefont{J.}~\bibnamefont{Sampaio}},
  \bibinfo{author}{\bibfnamefont{V.}~\bibnamefont{Cros}},
  \bibinfo{author}{\bibfnamefont{S.}~\bibnamefont{Rohart}},
  \bibinfo{author}{\bibfnamefont{A.}~\bibnamefont{Thiaville}},
  \bibnamefont{and} \bibinfo{author}{\bibfnamefont{A.}~\bibnamefont{Fert}},
  \bibinfo{journal}{Nature nanotechnology}  (\bibinfo{year}{2013}).

\end{thebibliography}


\end{document}


\title{Supplementary material: Dynamics and energetics of emergent magnetic monopoles in chiral magnets}
\author{Christoph Sch\"utte}
\affiliation{Institute for Theoretical Physics, University of Cologne, 50937 Cologne, Germany}
\author{Achim Rosch}
\affiliation{Institute for Theoretical Physics, University of Cologne, 50937 Cologne, Germany}

\begin{abstract} In this supplementary material we introduce  the Ginzburg Landau free energy, provide a discussion of the $T$ dependence of the core energy of the monopoles in various regimes, briefly investigate the creation rate of monopoles at the surface of the sample and provide technical details on  numerical implementations.
\end{abstract}
\date{\today}
 \maketitle
 
\section{Ginzburg-Landau theory}
Close to a phase transition (but still outside of the Ginzburg regime, where fluctuations dominate), one can describe the energetics of magnets by an effective Ginzburg Landau free energy.

After a rescaling of the coordinates, magnetization, magnetic field, and free energy, the Ginzburg Landau (GL) free energy density in the presence of the Dzyaloshinskii-Moriya interaction $\sim D \vec M \cdot ( \nabla \times \vec M)$ can be written in the following form \cite{muhlbauer2009skyrmion,bak1980theory}
\begin{equation}
F =(1+ t_0) \vec M^2 + (\nabla \vec M)^2 + 2 \vec M \cdot (\nabla \times \vec M) + \vec M^4 - \vec B \cdot \vec M
\label{eq:F}
\end{equation}
where $t_0$ measures the distance to the $B=0$ mean-field critical temperature.
 Here the length is measured in units where the pitch of the helix $\lambda$ (obtained for $B=0$) is given by $\lambda=2 \pi$ (further information on how the parameters of Eq.~(\ref{eq:F}) are related to microscopic parameters can be found in Ref. \cite{muhlbauer2009skyrmion}). In principal further contributions to the functional exist, but in the cubic B20 materials, in which skyrmions are observed, anisotropy terms can be neglected as they are higher order in spin-orbit coupling. The question whether dipolar interactions can be neglected is more subtle. Numerically, they give only very small contributions to the energetics of skyrmions (and therefore to the string tension) and the competing helical and conical phases in B20 magnets. As the core energy of the monopole is dominated by amplitude fluctuations and the magnetic exchange energy (see below) which are generically much stronger than dipolar interaction, they will give only small corrections to the core energy with the possible exception of a small region very close to the critical temperature ($T_c-T \ll \lambda_{SO}^2 T_c$, see below). For this reason, we neglect dipolar interactions which would also strongly increase the numerical cost of the simulations. 

\section{Energetics of monopoles and skyrmions: scaling analysis}

How the core energy of a monopole (MP) depends on temperature and microscopic parameters depends strongly on which regime is considered. Three length scales are of primary importance: (i) the underlying lattice spacing $a$, (ii) the typical length scale on which the {\em direction} of the magnetization changes due to spin orbit coupling,
which can be identified with the pitch of the helix $\lambda$ and, (iii) the length scale on which the {\em amplitude} of the magnetization changes, which can be identified with the radius of the core of the MP,  $R_c$. We always work in regimes where (as in the experiment \cite{milde2013unwinding}) the skyrmion radius is of order $\lambda$. Note that for single skyrmions embedded in ferromagnetic phases (we study only skyrmions in helical phases) this relation does not hold as $R_c$ depends strongly on the strength of the $B$ field in a ferromagnetic environment.

Due to the weakness of SO interactions, one always is in the regime $\lambda \gg a$. We will not discuss the regime $|t_0| \ll 1$ ($t_0<0$) where $R_c \sim \lambda$ (due to a fluctuation-induced first-order transition \cite{janoschek2013fluctuation}, this regime does probably also not exist experimentally) but focus on cases where $R_c \ll \lambda$ and $|t_0| \gg 1$. In more physical variables, the condition $|t_0|\gg 1$ translates to $T_c-T \gg \lambda_{SO}^2 T_c$, where $T_c$ is the mean-field transition temperature and $\lambda_{SO}$ is a dimensionless constant describing the strength of spin-orbit interaction (given by the ratio of Dzyaloshinskii Moriya interactions and exchange interactions \cite{muhlbauer2009skyrmion}). 
In this limit, the energetics of the MP core is not affected by the spin-orbit coupling but reflects the energy needed to suppress the {\em amplitude} of the magnetization in the core of the MP.  Two qualitatively different regimes have to be considered in this case.

{\it Regime 1: $\lambda_{SO}^2 T_c \ll T_c-T \ll T_c$:} In this regime, the GL theory, Eq.~(\ref{eq:F}), can be applied with $t_0<0$, $|t_0| \gg 1$ (but not too large, see below). Comparing the first and second term in
Eq.~(\ref{eq:F}) suggests that $R_c \sim \frac{1}{\sqrt{|t_0|}}\sim a \left(\frac{T_c}{T_c-T}\right)^{1/2}$. We have checked this statement using a more careful analysis based on the energetics of a hedgehog spin configuration, $\vec M(\vec r) = M(r) \hat{\Omega}(\theta,\phi)$, where $\hat{\Omega}(\theta,\phi)$ describes a spin-configuration winding once around the unit sphere and $M(r)$ describes variations of the amplitude of the magnetization on the length scale $R_c$. This ansatz is valid close to the MP core for $R_c \ll \lambda$. In this regime, the
core energy $E_c$ of the MP scales with $R_c^3 t_0 M^2\sim T_c  \left(\frac{T_c-T}{T_c}\right)^{1/2}$ with $M \sim \sqrt{T_c-T}$. In this regime, the string tension, $T_s$, i.e. the energy per length of a single skyrmion arises from changes of the direction of the magnetization and is given by the interplay of the second, third and last term of Eq. (\ref{eq:F}). For $B \lesssim M$ (the regime relevant for our study) it grows with $M^2$. As the energy density within the skyrmion is proportional to $1/\lambda^2$ and the skyrmion radius is given by $\lambda$, $T_s$ is independent of the strength of spin-orbit coupling
and we find $T_s \sim c_B  (T_c-T) / a $ with a prefactor $c_B$ of order $1$ which depends strongly on the ration of $B/M$ and is negative for small $B$ (positive for large $B$) as the skyrmion cost energy when embedded into the helical phase at low $B$. We therefore obtain
\begin{align}
\frac{R_c}{a} \sim \left(\frac{T_c}{T_c-T} \right)^{1/2}, \quad \frac{E_c}{T_c} \sim \left(\frac{T_c-T}{T_c} \right)^{1/2}, \quad \frac{T_s}{T_c/a} \sim c_{B}  \frac{T_c-T}{T_c} \quad \text{for }\  \lambda_{SO}^2 T_c \ll T_c-T \ll T_c \label{scalingR}
\end{align}

{\it Regime 2: $T \ll T_c$:} Upon lowering $T$, the magnetization grows and, ultimately, saturates. Also $R_c$ shrinks, saturating at a microscopic length scale typically given by the lattice constant (an exception are systems close to a quantum critical point). In simple models without frustration the two crossovers occur simultaneously at a temperature scale of the order of $T_c/2$. For temperatures small compared to this scale,
where $R_c \sim 1$ and the magnetization is saturated, also the core energy of the MPs saturate
at a value determined by microscopic parameters. It is dominated by the energy cost to have the magnetization on neighboring lattice sites to be aligned in a hedgehog configuration (instead of a parallel spin configuration). Generically, this will be an energy scale of the order of $T_c$.
We therefore conclude that in regime 2
\begin{align}
\frac{R_c}{a} \sim 1, \qquad \frac{E_c}{T_c} \sim 1, \qquad \frac{T_s}{T_c/a} \sim c_B  \qquad \text{for }\ T \ll T_c
\end{align}
fully consistent with Eq. (\ref{scalingR}).

It is important to note that in regime 1, the creation rate of MPs and antimonopoles (AMPs) $\propto e^{-E_c/T}$ is {\em not} exponentially suppressed. As our analysis of creation rates and the motion of MPs relied on the existence of only a small number of MPs, it is therefore not surprising that the analysis of the stochastic LLG equations was restricted to regime $2$. Indeed, for the highest temperature studied ($T=0.8$)  the time-averaged local magnetization far away from the MP core is of the order of $0.7$, still close to the saturation value of $1$ and in the distance of one lattice spacing from a MP core we find a magnetization of the order of $0.35$.

Note that the size of the core energies and string tensions obtained in the main text are fully consistent with the order-of-magnitude estimates given from the scaling analysis above.

\section{Energetics of monopoles close to the surface boundaries}

In this section we briefly comment on a surprising observation within our sLLG simulations not discussed in the main text: the creation rate of a single MPs at the surface of the simulated sample (we used open boundary condition) is within our simulations of a similar order of magnitude as the creation rate of MP-AMP pairs. This is unexpected, as naively the energy to create a MP-AMP pair should be approximately twice as large as the energy to create a single MP (or AMP). At least for the small samples used in our simulation, the creation rate at the boundary should therefore be much larger than the creation rate at the surface according to this hand-waving argument.

The argument misses, however, two effects: First, as discussed in the main text an effective attraction of MP-antimonole pairs at short distances effectively reduces the energy cost to create pairs in the bulk
(see Fig.~4 and 5 of the main text). A second effect is quantitatively of equal importance: the surface energy of the skyrmion.
Creating a MP at the surface implies that the skyrmion configuration at the surface is replaced by a helical configuration at the surface which is of higher energy. To quantify this extra energy, one can, for example
calculate for a cube of size $L$ with open boundary conditions the energy difference of the skyrmion and the helical phase, $\Delta E_S(L)$. The extra contribution to the surface energy arising from the skyrmion, $\Delta_S$ is calculated from $\Delta_S=\frac{\Delta E_S(L)-L T_S}{2}$, where $T_S$ is the energy per length of the skyrmion. For $t_0=-1.6$, for example, and $B=0.2$, we find $\Delta_S \approx 0.5$ while $E_c \approx 1.5$. Therefore, the creation of a MP-AMP pair in the bulk (energy $2 E_C \approx 3.0$ minus a correction from the binding energy) can be of similar size compared to the energy to create single MPs at the surface, $E_c+\Delta_S$.

\section{Numerical implementation}
\subsection{Numerical minimisation of the free energy}
The mean-field configurations of the magnetization can be studied by numerical minimization of an appropriately discretized Ginzburg-Landau functional $F[\vec M(\vec r)]$ either in position or momentum space. To study the energetics of the MP configuration we use a  discretization Eq.~(\ref{eq:F}) in real-space (we typically use $16$ spins per pitch of the helix, $\Delta x \approx 2 \pi/16$ and $50 \times 50 \times 50$ spins in total). Numerical minimization algorithms typically approach the closest local minimum and therefore the result depends on the initial magnetization configuration. In order to study mean-field configurations of systems which exhibit topological soliton solutions one has to choose an initial magnetization configuration which lies within the correct topological sector. For instance we would initialise a quadratic block of discretized spins with a helical configuration and a single Skyrmion embedded half way into the block which terminates with a magnetic MP at the centre of the block.

Unless the string tension $T_S$ vanishes, the minimisation algorithm will push the MP either upward or downward elongating or shortening the skyrmion in the process in order to minimize the free energy functional Eq.~(\ref{eq:F}). We pin the MP in place by enlarging the prefactor in front of the $\sim \vect M^2$ term in Eq.~(\ref{eq:F}) at the site where the MP's centre is located. Since the magnetization amplitude vanishes in the centre of the MP, for not too strong string tension $T_S$ and for sufficiently small $R_C$ this will hold the MP in place. If $R_C$ is much larger than the discretization length $a$ the variation of the magnetisation amplitude becomes smooth and the penalization of the magnetization is not sufficient to hold the MP in place.

For the numerical minimisation we used a generalisation of the conjugate gradient method as proposed by Fletcher and Reeves \cite{fletcher1964function} and as implemented in the GNU Scientific Library (GSL) \cite{galassi2006gnu}.

\subsection{Numerical integration of the stochastic Landau-Liftshitz-Gilbert equation}

For the numerical integration of the  stochastic Landau Lifshitz Gilbert equations (sLLG) \cite{garcia1998langevin}, see Eq. (1) of the main text, we use Heun's method for a  $35^3$ lattice with open boundary conditions in the $z$ and periodic boundary conditions in the $x$ and $y$ directions. The magnetic fields is applied in the $[\bar{1}\bar{1}2]$ direction, which is chosen perpendicular to the $[111]$ direction to avoid a tilting of helical phase which has an ordering vector in $[111]$ direction (due to the anisotropies of the discretisation the wave vector pins in this direction \cite{buhrandt2013skyrmion}). At $t=0$ we start from a configuration described by a helical spin state which has a single skyrmion embedded along the $[\bar{1}\bar{1}2]$ direction parallel to the magnetic field, see Fig.~1a of the main text, and use Heun's scheme with an appropriate discretisation of time ($\Delta t=0.01$) to numerically integrate the equation of motion, Eq.~(1) in the main text, - see the appendix of Ref.~\cite{garcia1998langevin} for more details. We calculate snapshots of the magnetic configuration averaged over short times ($\bar{t}=1$) to remove most of the spurious MP-AMP pairs due to single thermal spin flips. The winding number in each discretization cube of the snapshot can be determined by the calculation of the solid angle for each triangle in the triangulation of the cube. The solid angle of 3 spins at the vertices of a triangle is given by the Oosterom and Strackee algorithm \cite{van1983solid}
\begin{equation}
\tan\left( \frac{\Omega}{2} \right) = \frac{\hat{n}_1 \cdot ( \hat{n}_2 \times \hat{n}_3)}{1 + \hat{n}_1 \cdot \hat{n}_2 + \hat{n}_2 \cdot \hat{n}_3 + \hat{n}_3 \cdot \hat{n}_1}
\end{equation}
where $\Omega$ is the solid angle and $\hat{n}_i$ is the spin direction at vertex $i$. In this way the location of each magnetic MP in the system can be determined. We use a tracking algorithm to connect the MP position in subsequent snapshots into MP trajectories. It becomes exceedingly difficult to program a reliable tracking algorithm for situations close to $T_C$ where an increasing number of thermal MP-AMP pairs are created in the bulk. This limits the accessible parameter regime to regime 2, see above. 

\bibliographystyle{apsrev}
\bibliography{bibliography}